\documentclass[prb,preprint]{revtex4}

\pdfoutput=1

\usepackage{graphicx}
\usepackage{latexsym}
\usepackage{MnSymbol}

\begin{document}

\title{Retrograde doping dependence of charge order in $\mathbf{La_{1.8-x}Eu_{0.2}Sr_xCuO_4}$}

\medskip 

\date{November 7, 2021} \bigskip

\author{Manfred Bucher \\}
\affiliation{\text{\textnormal{Physics Department, California State University,}} \textnormal{Fresno,}
\textnormal{Fresno, California 93740-8031} \\}

\begin{abstract}
Comparison with $La_{2-z-x}Nd_zSr_xCuO_4$ ($z=0,\;0.4)$ shows that the
retrograde doping dependence of charge order in $La_{1.8-x}Eu_{0.2}Sr_xCuO_4$ must be caused by a mechanism transcending the common charge-order generation in the hole-doped lanthanum cuprates. This could be a bond-stretching phonon with comparable momentum, $q \approx 0.24$ r.l.u., as observed in $La_{1.675}Eu_{0.2}Sr_{0.125}CuO_4$.

\end{abstract}

\maketitle
\section{CHARGE ORDER IN HOLE-DOPED LANTHANUM CUPRATES}

The study of charge-order (CO) and magnetization (M) stripes is a valuable tool in the investigation of the pseudogap phase of high-$T_c$ superconducting copper oxides. In the family of the lanthanum cuprates (`214'), doped with alkaline-earth $Ae = Sr, Ba$, and possibly co-doped with lanthanides $Ln = Nd, Eu$, 
the dependence of their incommensurability on $Ae$-doping $x$ is related as
\begin{equation}
    q_{CO}^{CuO_2}(x) = 2 q_{M}^{CuO_2}(x) \;,
\end{equation}
in reciprocal lattice units (r.l.u.).
Recent experiments\cite{1,2,3} on the compounds $La_{2-x}Sr_xCuO_4$ and $La_{1.6-x}Nd_{0.4}Sr_xCuO_4$  have explored stripes in an extended doping range range, 
$0.12 < x \le 0.21$ and $0.26$, respectively.
They confirmed that their incommensurability increases monotonically, 
\begin{equation}
q_{CO}^{CuO_2}(x)= \frac{\Omega^{\pm}}{{2}}\sqrt {x - \tilde{p}}\;,\;\;\;\;\;x < \hat{x}  \; ,\cite{4}
\end{equation}
 but levels off, beyond a ``watershed'' concentration $\hat{x}$ (depending on the dopant and co-dopant species),
to a constant plateau,
\begin{equation}
    q_{CO}^{CuO_2}(x) = \frac{\sqrt{2}}{2} \sqrt{\hat{x} - \tilde{p}} \;,\;\;\;\; x \ge \hat{x} \;,
\end{equation}
(see Fig. 1). Such a levelling-off had been observed a long time ago,\cite{5}
but was recently explained\cite{6} in terms of overflow of doped holes from saturated $CuO_2$ planes to the bracketing $La_{1-z}Nd_zO$ ($z = 0, \; 0.2$) layers with incommensurability
\begin{equation}
    q_{CO}^{LaO}(x)  = \frac{\sqrt{2}}{2}\sqrt {\frac{x - \hat{x}}{2} }\;,\;\;\;\;\;  x \ge \hat{x} \; .
\end{equation}
Specifically, those experiments\cite{1,2,3} determined the watershed values $\hat{x}(La_{2-x}Sr_xCuO_4)=0.125$ and $\hat{x}(La_{1.6-x}Nd_{0.4}Sr_xCuO_4)=0.17$ (see Fig. 1).

A related issue is the closing of the pseudogap at $T=0$, that is, the doping $x^*$ (quantum critical point) where the pseudogap temperature vanishes, $T^*(x^*)=0$. 
As shown in Fig. 2, 
\linebreak
$T^*(x)$ data  of $La_{2-x}Sr_xCuO_4$ and co-doped $La_{2-z-x}Ln_zSr_xCuO_4$ ($Ln = Nd, Eu;\; z = 0.4$, $0.2$), 
obtained from transport (Hall effect, Nernst effect) and spectroscopic (ARPES) measurements,\cite{8} fall on a common, linearly decreasing straight line until rapidly dropping off to $T^*=0$ at $x^* = 0.18$ for $La_{2-x}Sr_xCuO_4$, but continuing commonly for $La_{2-z-x}Ln_zSr_xCuO_4$ ($Ln = Nd, Eu$) to $x=0.21$, then for only $La_{1.6-x}Nd_{0.4}Sr_xCuO_4$,
until a drop-off at $x^*=0.23$. The latter finding indicates a close similarity as a result of $Ln = Nd, Eu$ co-doping.

\includegraphics[width=6in]{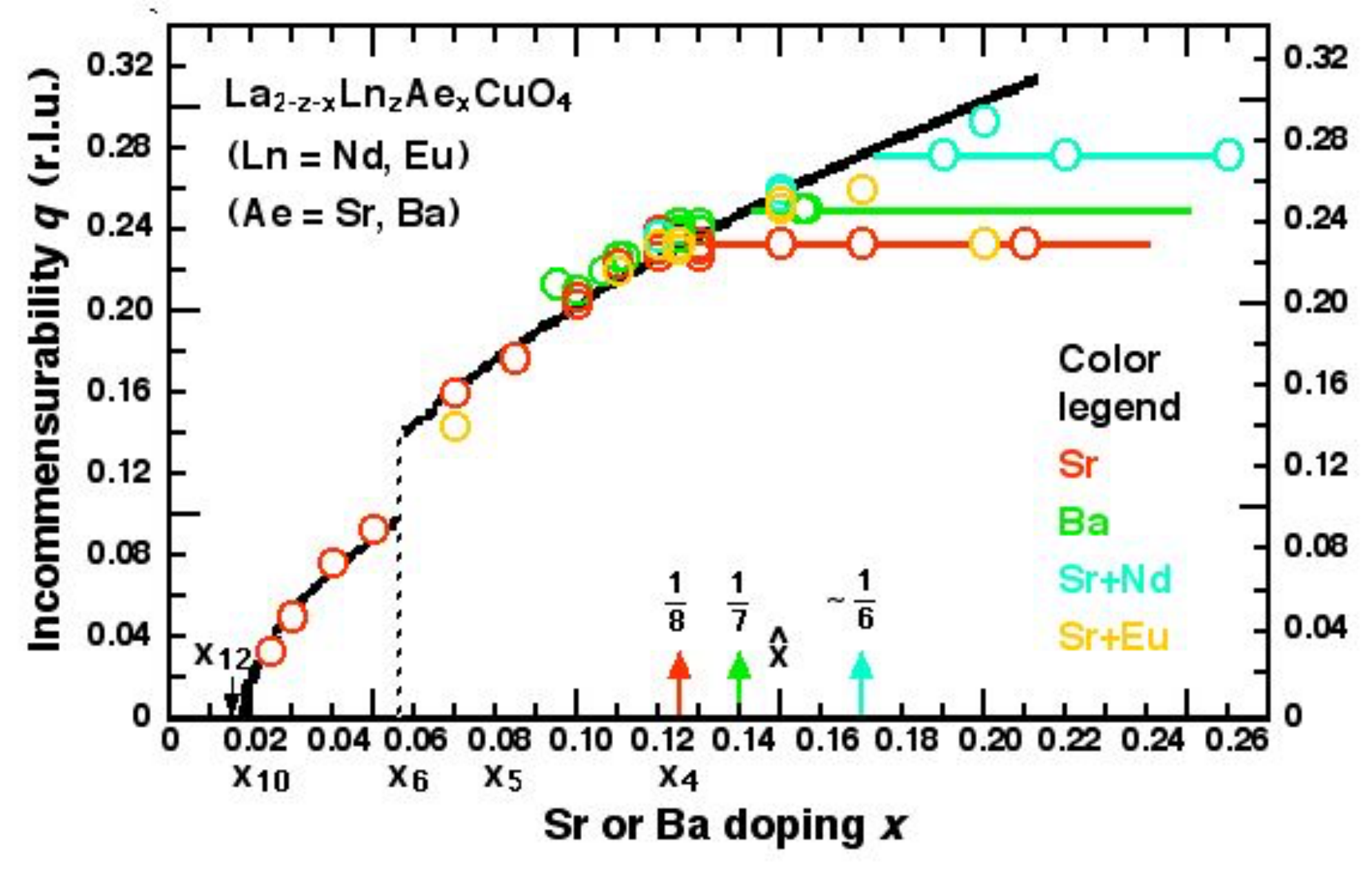}  \footnotesize 

\noindent FIG. 1. Incommensurability of charge-order stripes, $q(x) = q_{CO}^{CuO_2}(x)$, and of magnetization stripes, $q(x) = 2q_M^{CuO_2}(x)$, in $La_{2-z-x}Ln_zAe_{x}CuO_{4}$ ($Ln = Nd, Eu; z = 0, 0.4, 0.2$) due to doping with $Ae = Sr$ or $Ba$. Circles show data from X-ray diffraction or neutron scattering (cf. Ref. 6). 
The broken solid curve is a graph of Eq. (2), calculated with a constant offset value, $\tilde{p} = 0.02$. 
Commensurate doping concentrations are denoted by $x_n \equiv 2/n^2$.
The discontinuity at $x_6 \simeq 0.056$ is caused by a change of stripe orientation, relative to the planar crystal axes, from diagonal for $x<x_6$ to parallel for $x>x_6$. 
The curve holds for temperature near $T=0$ and is accurate for low doping, $x<0.09$. Neglect of the doping dependence of the offset value, $\tilde{p}(x) < 0.02$, causes the slight deviation of the curve (too low) from most data in the doping range $x > 0.09$. 
Doping beyond watershed concentrations, $\hat{x}_{Sr} = 0.125 = 1/8$,  $\hat{x}_{Ba} = 0.14 \simeq 1/7$ and $\hat{x}_{Sr+Nd} = 0.17 \approx 1/6$, yields  constant stripe incommensurabilities,
$q_c(x) = 0.235$ ($Sr$), 0.25 ($Ba$) and 0.278 ($Sr$+$Nd$), given by Eq. (3) (horizontal lines). \normalsize

\section{CHARGE ORDER IN $\mathbf{La_{1.8-x}Eu_{0.2}Sr_xCuO_4}$}

Very recently neutron scattering and resonant soft X-ray scattering experiments\cite{9} have provided incommensurability data of $La_{1.8-x}Eu_{0.2}Sr_xCuO_4$ in the doping range $0.07 \le x \le 0.20$. 
Only the low-temperature data ($T = 24$ K) are considered in this note. They are included in Fig. 1. The values for $0.07 \le x \le 0.15$ are in good agreement with previous $q_{CO}^{CuO_2}(x)$ data of $La_{1.8-x}Eu_{0.2}Sr_xCuO_4$. The value at $x=0.15$ follows the square-root curve  of Eq. (2) beyond the watershed concentration $\hat{x} = 0.125$ of $La_{2-x}Sr_xCuO_4$. From the close similarity of the $T^*(x)$ data of the two $La_{2-z-x}Ln_zSr_xCuO_4$ compounds, $Ln = Nd, Eu$, one would expect a very close levelling-off for both cases of co-doping, that is, at $\hat{x} = 0.17$ to $q_{CO}^{CuO_2}= 0.278$. 
Surprisingly, this is not borne out by the novel data of $La_{1.8-x}Eu_{0.2}Sr_xCuO_4$ at $x = 0.17$ and $x=0.20$. The possibility of a watershed value $\hat{x} = 0.154$ of $La_{1.8-x}Eu_{0.2}Sr_xCuO_4$ that would determine a constant level, coinciding with $q_{CO}^{CuO_2}(0.17) = 0.259$, must be ruled out because of the lower value $q_{CO}^{CuO_2}(0.20) = 0.234$. What could be the reason for the retrograde incommensurability beyond $x=0.17$?

The monotonic doping dependence of charge-order incommensurability on $Sr$-doping $x$, Eqs. (2) and (4), is the result of Coulomb repulsion between doped holes, residing pairwise at $O$ atoms at anion lattice sites.\cite{6} Apart from the $CuO_2$ planes and $La_{0.9}Eu_{0.1}O$ layers, there are no other places to which $O$ atoms could escape. Beyond $\hat{x}$, the hole density in the $CuO_2$ planes and $La_{0.9}Eu_{0.1}O$ layers must be in balance. This eliminates the possibility of removing holes from the $CuO_2$ planes by adding them to the $La_{0.9}Eu_{0.1}O$ layers. Where, then, did the doped holes, lost in the transition $x=0.17 \rightarrow 0.20$, go? Their ``disappearance'' transcends the mechanism of charge-order generation. What could that be?

\includegraphics[width=5.8in]{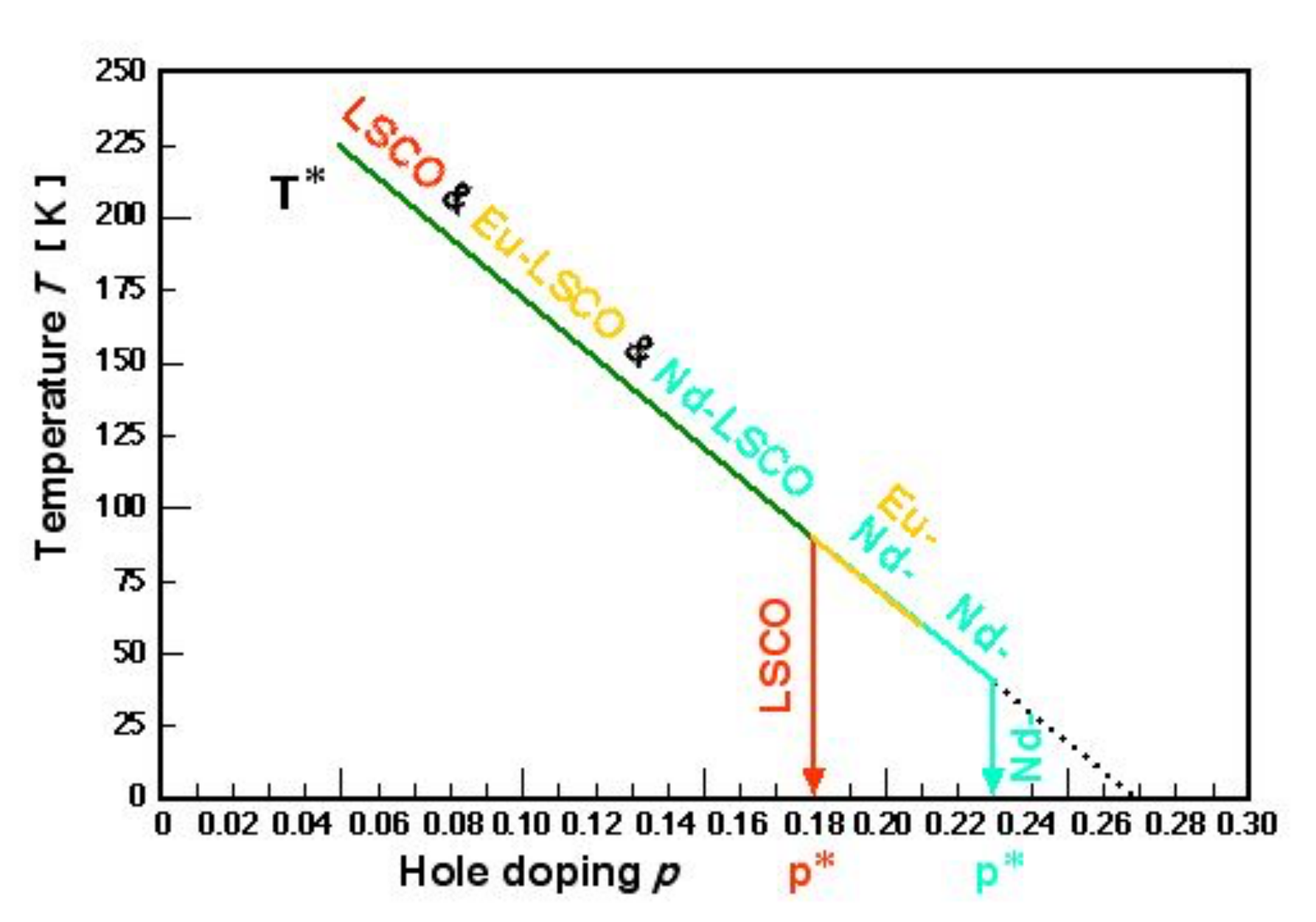}  \footnotesize 

\noindent FIG. 2. Doping dependence of the pseudogap temperature $T^*$ of $La_{2-z-x}Ln_zSr_xCuO_4$ ($Ln = Eu, Sr;
\linebreak z = 0, \;0.2,\; 0.4$) and quantum critical points $p^*$ where the pseudogap closes (graph simplified after Ref. 8).  \normalsize

\section{BOND-STRETCHING PHONONS}

Resonant inelastic X-ray scattering (RIXS) in $La_{1.675}Eu_{0.2}Sr_{0.125}CuO_4$ has revealed bond-stretching phonons of momentum $q \approx 0.24$ in the $CuO_2$ plane, and corresponding phonon softening, that significantly weakens charge order of comparable incommensurability.\cite{10}
In oscillating bond-stretching, $O^{2-}$ ions, bracketing axially next-nearest $Cu^{2+}$ neighbors, move in a breathing mode.
It can be assumed that such phonons are present in $La_{1.8-x}Eu_{0.2}Sr_xCuO_4$ at other doping levels, too, including $x=0.17$ and $x=0.20$ with incommensurabilities $q_{CO}^{CuO_2} \approx 0.24$. The double-hole-bearing $O$ atoms of those compounds will likewise participate in the breathing mode. This effectively reduces their (average) presence at anion lattice sites. In this sense they are \textit{partially} lost to the hole density that generates charge order.
Qualitatively, the reduced concentration of $O$ atoms at anion lattice sites causes a widening of their superlattice spacing with consequent decrease of $q_{CO}^{CuO_2}(x)$ incommensurability.

A view of Fig. 2 shows that the  quantum critical point $p^*$ of $La_{1.8-x}Eu_{0.2}Sr_xCuO_4$ has not been determined.
It remains an open question how the retrograde incommensurabily of this compound affects the closing of its pseudogap.

Are bond-stretching phonons present in $Nd$ co-doped $La_{1.6-x}Nd_{0.4}Sr_xCuO_4$? None have been reported.
Their absence would explain why no retrograde charge order is observed
in that compound.



\begin{thebibliography}{20}
\bibitem{4} The  stripe-orientation factor in Eq. (2) is $\Omega^{+}=\sqrt{2}$ for $x > x_6 = 2/6^2  \simeq 0.056$ when stripes are parallel to the $a$ or $b$ axis, but $\Omega^{-} = 1$ for $x < x_6$ when stripes are diagonal. The offset value $\tilde{p}$ under the radical is the hole concentration necessary to keep three-dimensional antiferromagnetism (3D-AFM) suppressed.
\end{thebibliography}
\end{document}